\documentclass[final,3p,times,onecolumn]{elsarticle}
\usepackage{amsmath}
\usepackage{amssymb}
\usepackage{graphicx}

\journal{Physics Letters B}

\newcommand{\beq}{\begin{equation}}
\newcommand{\eeq}{\end{equation}}
\newcommand{\beqn}{\begin{eqnarray}}
\newcommand{\eeqn}{\end{eqnarray}}

\newcommand{\n}{\mathbf{n}}
\newcommand{\p}{\mathbf{p}}

\newcommand{\si}{\mbox{{\boldmath$\sigma$}}}

\newcommand{\al}{\mbox{${\alpha}$}}

\newcommand{\ga}{\mbox{${\gamma}$}}

\newcommand{\de}{\mbox{${\delta}$}}

\newcommand{\ep}{\mbox{${\varepsilon}$}}
\newcommand{\om}{\mbox{${\omega}$}}

\newcommand{\na}{\mbox{${\nabla}$}}
\newcommand{\pa}{\mbox{${\partial}$}}

\begin{document}

\begin{frontmatter}

\title{Gravitational four-fermion interaction on the Planck scale}

\author{I.B.\,Khriplovich\footnote{khriplovich@inp.nsk.su}\\
Budker Institute of Nuclear Physics\\
630090 Novosibirsk, Russia}

\begin{abstract}

The four-fermion gravitational interaction is induced by torsion,
and gets essential on the Planck scale. On this scale, the
axial-axial contribution dominates strongly the discussed
interaction. The energy-momentum tensor, generated by this
contribution, is analyzed, as well as stability of the problem
with respect to compression. The trace of this energy-momentum
tensor can be negative.

\end{abstract}

\end{frontmatter}

1. The observation that, in the presence of torsion, the interaction
of fermions with gravity results in the four-fermion interaction
of axial currents, goes back at least to \cite{ki}.

We start our discussion of the four-fermion gravitational
interaction with the analysis of its most general form.

As has been demonstrated in~\cite{hol}, the common action for the
gravitational field can be generalized as follows:
\beq\label{sg}
\noindent S_g = -\frac{1}{16\pi G}\int d^4 x(-e)e^{\mu}_I
e^{\nu}_J\left( R^{IJ}_{\mu\nu}
-\frac{1}{\ga}\tilde{R}^{IJ}_{\mu\nu}\right);
\eeq
\noindent here and below $G$ is the Newton gravitational constant,
$I,J=0,1,2,3$ are internal Lorentz indices, $\mu,\nu=0,1,2,3$ are
space-time indices, $e_\mu^I$ is the tetrad field, $e$ is its
determinant, and $e^{\mu}_I$ is the object inverse to $e_\mu^I$.
The curvature tensor is
\[
R^{IJ}_{\mu\nu}=-\pa_{\mu} \om^{IJ}{}_\nu+\pa_{\nu}\om^{IJ}{}_\mu+
\om^{IK}{}_{\mu}\om_K {}^{J}{}_\nu -\om^{IK}{}_\nu\om_K{}^J{}_\mu,
\]
here $\om^{IJ}_{\mu}$ is the connection. The first term in
(\ref{sg}) is in fact the common action of the gravitational field
written in tetrad components.

The second term in (\ref{sg}), that with the dual curvature tensor
\[
\tilde{R}^{IJ}_{\mu\nu}=
\,\frac{1}{2}\,\varepsilon^{IJ}_{KL}R^{KL}_{\mu\nu}\,,
\]
does not vanish in the presence of spinning particles
generating torsion \cite{pero}.

As to the so-called Barbero-Immirzi parameter $\ga$, its numerical
value
\beq
\ga = 0.274
\eeq
was obtained for the first time in \cite{kk}, as the solution of
the "secular" equation
\beq
\sum_{j=1/2}^{\infty} (2j+1) e^{- 2\pi\gamma \sqrt{j(j+1)}} = 1,
\eeq
derived in the mentioned paper.

Thus derived effective four-fermion interaction of axial currents
is \cite{pero}:
\beq\label{sa}
S_A= \,\frac{3}{2}\,\pi\, \frac{\ga^2}{\ga^2 + 1}\,\,G \int
d^4x\;(-e)\; \,\eta_{IJ}A^I A^J.
\eeq
Here and below $\eta_{IJ} = {\rm diag}(1,-1,-1,-1)$\,, $A^I$ is
the net fermion axial current:
\beq\label{A}
A^I = \sum_a A_a^I = \sum_a \bar{\psi}_a\,\ga^5\,\ga^I\,\psi_a\,;
\eeq
the sum extends over all sorts of fermions with spin 1/2. This
result (\ref{sa}) corresponds (up to a factor) to that derived long ago
in~\cite{ki}.

In the absence of the pseudoscalar term in gravitational
action (\ref{sg}), i.e. for $\ga \to \infty$, expression
(\ref{sa}) simplifies to
\beq\label{sa1}
S_A^{\infty} = \,\frac{3}{2}\,\pi\,G \int d^4x\;(-e)\;
\,\eta_{IJ}A^I A^J.
\eeq
This effective gravitational four-fermion interaction in the limit
$\ga \to \infty$ was derived previously in \cite{ker} (when
comparing the corresponding result from \cite{ker} with
(\ref{sa1}), one should keep in mind that the convention
$\eta_{IJ} = {\rm diag}(-1,1,1,1)$, used in \cite{ker}, differs in
sign from ours).

Two other contributions to the effective gravitational
four-fermion interaction (\ref{sa}) arise as follows. The common
action for fermions in gravitational field
\beq
S_f = \int d^4 x\,(-e)\,\frac{1}{2}[
 \,\bar{\psi}\,\ga^I\,e^{\mu}_I\,i\na_\mu \psi
 - \,i\,
 \overline{\na_\mu\psi}\,\ga^{I}e_{I}^\mu\psi]
\eeq
can be generalized in the following way (see \cite{fmt}):
\begin{eqnarray}\label{sf}
S_f &=& \int d^4 x \,(-e)\, \frac{1}{2}[
 (1-i\al)\,\bar{\psi}\,\ga^I\,e^{\mu}_I\,i\na_\mu \psi
 - (1+i\al)\,i\,
 \overline{\na_\mu\psi}\,\ga^{I}e_{I}^\mu\psi];
\end{eqnarray}
here
 \[
 \na_\mu = \pa_\mu -
 \frac{1}{4}\,\om^{IJ}_{\;\;\;\;\mu}\,\ga_I\ga_J\,,\quad
 [\na_\mu,\na_\nu] = \frac{1}{4}\,R^{IJ}_{\mu\nu}\,\ga_I\ga_J\,.
 \]
The real constant $\al$ introduced in (\ref{sf}) is of no
consequence, generating only a total derivative, if the theory is
torsion free. However, in the presence of torsion this constant
gets operative. In particular, as demonstrated in \cite{fmt}, it
results in the following generalization of the four-fermion
interaction (\ref{A}):
\beq\label{ff}
S_{ff}= \,\frac{3}{2}\,\pi\,\,G\frac{\ga^2}{\ga^2 + 1}\, \int
d^4x\,(-e)\; \left[\,\eta_{IJ}A^I A^J
+ \,\frac{\al}{\ga}\;\eta_{IJ}(V^I A^J + A^I V^J) -\al^2\,
\eta_{IJ}V^I V^J\,\right];
\eeq
here $V^I$ is the net fermion vector current:
\beq\label{V}
V^I = \sum_a V_a^I = \sum_a \bar{\psi}_a\,\ga^I\,\psi_a\,,
\eeq
and this sum, as well as the analogous one (\ref{A}) for the axial
current, extends over all sorts of elementary fermions with spin
1/2; of course, both $A^I_a$ and $V^I_a$ are neutral currents. As
to the negative sign at the $AA$ term in the corresponding formula
of \cite{fmt}, it is incorrect (I am grateful to A.A. Pomeransky
for pointing out this fact to me).

In fact, $VA$ and $VV$ terms in formula (\ref{ff}) are of no real
interest for the problem considered. Indeed, as follows from
simple dimensional arguments, interaction (\ref{ff}), being
proportional to the Newton constant $G$ and to the particle number
density squared, gets essential and dominates over the common
interactions only at very high densities and temperatures, i.e. on
the Planck scale and below it. Under these extreme conditions, the
number densities of both fermions and antifermions increase, due
to the pair creation, but the total vector current density $V^I$
remains intact. However, the situation with the axial current
density $A^I$ is quite different. As distinct from the $C$-odd
vector current $V^I$, the axial one $A^I$ is $C$-even. Therefore,
fermions and antifermions contribute to $A^I$ with the same sign,
so that $A^I$ grows together with density and temperature. Thus,
if we confine to the Planckian regime (the only one, where the
four-fermion gravitational interaction (\ref{ff}) can be
essential), the $V$-\,dependent contributions to this interaction
can be neglected, and interaction (\ref{ff}) reduces to the
axial-axial one (\ref{sa}).

Thus, the $V$-\,dependent contributions to the four-fermion
gravitational interaction (\ref{ff}), i.e. terms proportional to
$\al$ and $\al^2$, are negligibly small everywhere: above the
Planck scale they are small together with the purely axial $AA$
interaction, on the Planck scale and below it they are small as
compared to the $AA$ interaction.

\vspace{3mm}

2. Let us consider now the energy-momentum tensor (EMT)
$T_{\mu\nu}$ generated by action (\ref{sa}). Therein, the scalar
product $\eta_{IJ}A^I A^J$ has no explicit dependence at all
either on the metric tensor, or on its derivatives. The metric
tensor enters action $S_A$ only via $\;-e = \sqrt{-g}$, so that
\beq\label{sat}
S_A= \,\frac{3}{2}\,\pi\, \frac{\ga^2}{\ga^2 + 1}\,\,G \int
d^4x\;\sqrt{-g}\; \,\eta_{IJ}A^I A^J.
\eeq
Since $S_A$ is independent of the derivatives of metric tensor,
the corresponding EMT is given by relation
\beq
\frac{1}{2}\;\sqrt{-g}\;T_{\mu\nu} = \frac{\de}{\de g_{\mu\nu}}
S_A\,.
\eeq
Now, with identity
\beq
\frac{1}{\sqrt{-g}}\frac{\partial\sqrt{-g}}{\partial g^{\mu\nu}} =
- \frac{1}{2}\, g_{\mu\nu}\,,
\eeq
we arrive at the following expression for the EMT:
\[
T_{\mu\nu} = - \frac{3\pi}{2} \frac{\ga^2}{\ga^2 + 1}\,
g_{\mu\nu}\,\eta_{IJ}A^I A^J\,,
\]
or, in the tetrad components,
\beq\label{emt}
T_{MN} = - \frac{3\pi}{2} \frac{\ga^2}{\ga^2 + 1}\,
\eta_{MN}\,\eta_{IJ}A^I A^J\,.
\eeq
We note first of all that this EMT corresponds to the equation of state
\beq
p = -\ep;
\eeq
here and below $\ep = T_{00}$ is the energy density, and $p = T_{11} = T_{22} = T_{33}$ is the pressure.

Let us analyze this expression for the interaction of two
ultrarelativistic fermions (labelled $a$ and $b$) in their
center-of-mass system.

The axial current of fermion $a$ (both of particle and
antiparticle!) is
\[
A^I_a = \frac{1}{4E^2}\;\phi^\dagger_a\{E\, \si_a(\p^\prime +
\p),\;(E^2  - (\p^\prime \p))\si_a
+ \p^\prime(\si_a\p)
+ \p(\si_a \p^\prime) - i [\p^\prime \times \p]\}\phi_a =
\]
\beq
= \frac{1}{4}\;\phi^\dagger_a\{\si_a(\n^\prime + \n),\;(1 -
(\n^\prime \n))\si_a
 + \n^\prime(\si_a\n) + \n(\si_a \n^\prime) - i [\n^\prime \times \n]\}\phi_a;
\eeq
here $E$ is the energy of fermion $a$, $\n$ and $\n^\prime$ are
the unit vectors of its initial and final momenta $\p$ and
$\p^\prime$, respectively; of course, under the discussed extreme
conditions all fermion masses can be certainly neglected. In the
center-of-mass system, the axial current of fermion $b$ is
obtained from this expression by changing the signs: $\n \to -\n$,
$\n^\prime \to -\n^\prime$. Then, after averaging over the
directions of $\n$ and $\n^\prime$, we arrive at the following
semiclassical expressions for the nonvanishing components of the
energy-momentum tensor, i.e. for the energy density $\ep$ and
pressure $p$ (for the correspondence between $\ep$, $p$ and EMT
components see \cite{ll}, \S\;35):
\beq\label{ep}
\ep = -\,\frac{\pi}{48}\,\frac{\ga^2}{\ga^2 +
1}\;G\,\sum_{a,\;b}\rho_a \,\rho_b\,(3 - 11\,<\si_a\si_b>)
=\,-\,\frac{\pi}{48}\,\frac{\ga^2}{\ga^2 + 1}\;G\,\rho^2\,(3 -
11\,\zeta)\,;
\eeq
\beq\label{p}
p = \frac{\pi}{48}\,\frac{\ga^2}{\ga^2 + 1}\;G
\,\sum_{a,\;b}\rho_a\,\rho_b\,(3 - 11\,<\si_a\si_b>)
=\frac{\pi}{48}\,\frac{\ga^2}{\ga^2 + 1}\;G\,\rho^2\,(3 -
11\,\zeta)\,;
\eeq
here and below $\rho_a$ and $\rho_b$ are the number densities of
the corresponding sorts of fermions and antifermions, $\rho =
\sum_a \rho_a$ is the total density of fermions and antifermions,
the summation $\sum_{a,\,b}$ is performed over all sorts of
fermions and antifermions; $\zeta = \,<\si_a\si_b>$ is the average
value of the product of corresponding $\si$-matrices, presumably
universal for any $a$ and $b$. Since the number of sorts of
fermions and antifermions is large, one can neglect here for
numerical reasons the contributions of exchange and annihilation
diagrams, as well as the fact that if $\si_a$ and $\si_b$ refer to
the same particle, $<\si_a\si_b> = 3$. The parameter $\zeta$\,,
just by its physical meaning, in principle can vary in the
interval from 0 (which corresponds to the complete thermal
incoherence or to the antiferromagnetic ordering) to 1 (which
corresponds to the complete ferromagnetic ordering).

Let us note that, according to (\ref{ep}), the contribution of the
gravitational spin-spin interaction to energy density is positive,
i.e. the discussed interaction is repulsive for fermions with
aligned spins. This our conclusion agrees with that made long ago
in \cite{ker}.

To simplify the further discussion, we will confine to the region
somewhat below the Planck scale, so that one can neglect effects
due to the common fermionic EMT, originating from the Dirac
Lagrangian and linear in the particle density $\rho$.

\vspace{3mm}

3. A reasonable dimensional estimate for the temperature $\tau$ of
the discussed medium is
\beq\label{e3}
\tau \sim \,\rho^{1/3}\,\sim \,m_{\rm {Pl}}\
\eeq
(here and below $m_{\rm {Pl}}$ is the Planck mass). This
temperature is roughly on the same order of magnitude as the
energy scale $\om$ of the discussed interaction
\beq\label{om}
\om \sim \,G\,\rho\,\sim \,m_{\rm {Pl}}\,.
\eeq
Numerically, however, $\tau$ and $\om$ can differ essentially.
Both options, $\tau > \om$ and $\tau < \om$, are conceivable.

If the temperature is sufficiently high, $\tau \gg \om$, it destroys the spin-spin
correlations in formulas (\ref{ep}) and (\ref{p}). In the opposite
limit, when $\tau \ll \om$, the energy density
(\ref{ep}) is minimized by the antiferromagnetic spin ordering.
Thus, in both these limiting cases the energy density and pressure
simplify to
\beq\label{ep0}
\ep =  -\,\frac{\pi}{16}\,\frac{\ga^2}{\ga^2 + 1}\;G\,\rho^2;
\eeq
\beq\label{p0}
p = \frac{\pi}{16}\,\frac{\ga^2}{\ga^2 + 1}\;G \, \rho^2\,.
\eeq

\noindent The energy density $\ep$, being negative and proportional to
$\rho^2$, decreases with the growth of $\rho$. On
the other hand, the common positive pressure $p$ grows together
with $\rho$. Both these effects result in the compression of the
fermionic matter, and thus make the discussed system unstable.

A curious phenomenon can be possible if initially the temperature
is sufficiently small, $\tau < \om$, so that equations (\ref{ep0}), (\ref{p0})
hold. Then the matter starts compressing, its temperature grows,
and the correlator $\zeta = \,<\si_a\si_b>$ arises. When (and if)
$\zeta$ exceeds its critical value $\zeta_{cr} = 3/11$, the
compression changes to expansion. Thus, we arrive in this case at the big bounce situation.

\vspace{3mm}

4. The last remark is as follows. It is well-known that for a
system of point-like particles with the electromagnetic
interaction among them, the trace $T^\mu_\mu$ of its EMT satisfies
the condition
\beq\label{tr1}
T^\mu_\mu = g_{\mu\nu}T^{\mu\nu} \geq 0\,.
\eeq
The assumption is usually made that this condition is valid also
for other interactions existent in Nature (see in this connection
\cite{ll}, \S\;34).

However, in our problem the trace
\beq T^\mu_\mu = \ep - 3p
= -\,\frac{\pi}{12}\,\frac{\ga^2}{\ga^2 + 1}\;G\,\rho^2\,(3 - 11\,
\zeta)
\eeq
is negative if $\zeta < 3/11$.

This feature of our problem is closely related to the fact that
the system under discussion is unstable with respect to
compression for $\zeta < 3/11$.

\begin{center}***\end{center}

I truly appreciate numerous helpful discussions with A.A.
Pomeransky. I am grateful also to D.I.~Diakonov, V.F.~Dmitriev,
A.D.~Dolgov, A.S.~Rudenko, and V.V.~Sokolov for their interest to
the work and useful discussions.

The investigation was supported in part by the Russian Ministry of Science, by the Foundation
for Basic Research through Grant No. 11-02-00792-a, by the Federal
Program "Personnel of Innovational Russia" through Grant No.
14.740.11.0082, and by the Grant of the Government of Russian
Federation, No. 11.G34.31.0047.

\renewcommand{\refname}

\end{document}